\documentclass[12pt]{iopart}

\usepackage{color}
\usepackage{graphicx}
\usepackage{iopams}

\begin{document}

\title{Majorana zero modes and long range edge correlation in interacting Kitaev chains: analytic solutions and density-matrix-renormalization-group study}

\author{Jian-Jian Miao$^{1,2,3}$, Hui-Ke Jin$^{1,2}$, Fu-Chun Zhang$^{3,1,2}$ and Yi Zhou$^{1,2}$}
\address{$^1$ Department of Physics, Zhejiang University, Hangzhou 310027, China}
\address{$^2$ Collaborative Innovation Center of Advanced Microstructures, Nanjing 210093,
China}
\address{$^3$ Kavli Institute for Theoretical Sciences, University of Chinese Academy of Sciences, Beijing 100190, China}
\ead{yizhou@zju.edu.cn}
\pacno{
 71.10.Fd,
 71.10.Pm,
 74.20.-z,
 03.75.Lm
}

\begin{abstract}
We study Kitaev model in one-dimension with open boundary condition by using exact analytic methods for non-interacting system at zero chemical potential as well as in the symmetric case of $\Delta=t$,
and by using density-matrix-renormalization-group method for interacting system with nearest neighbor repulsion interaction.
We suggest and examine an edge correlation function of Majorana fermions to characterize the long range order in the topological superconducting states and study the phase diagram of the interating Kitaev chain.
\end{abstract}

\maketitle

\section{Introduction}

Majorana\cite{Majorana} zero mode (MZM) has attracted a lot of attention in the recent years\cite{Wilczek,Beenakker,Elliott}, which may emerge as a novel excitation in some topological condensed matter systems.
MZMs obey non-Abelian statistics and have potential application to build robust qubits against decoherence in quantum computation\cite{Nayak2008,Alicea}.
The emergence of MZMs has been theoretically proposed in a number of condensed matter systems, including chiral $p$-wave superconductors\cite{Kitaev,DasSarma2006}, $\nu=5/2$ fractional quantum Hall system\cite{Moore1991},
the interface between a topological insulator and an $s$-wave superconductor\cite{Fu2008},
proximity-induced superconductor for spin-orbit coupled nanowires\cite{Oreg2010,Lutchyn2010a},
spin-orbit coupled semiconductor with externally applied Zeeman field\cite{Lutchyn2010b,Tewari2010,Alicea2010}, and ferromagnetic atoms in proximity to superconductors\cite{Beenakker2011,Yazdani2013}.
There also exist various experimental efforts to realize and detect MZMs in these proposed systems\cite{Mourik,Rokhinson,Das,Deng,Churchill,Lee,Yazdani,Jia12,Jia15,Jia16}.

Among these candidates, the one-dimensional (1D) systems are of special theoretical interest for possible generalization to interacting systems.
The interaction may change properties drastically in 1D systems.  The Fermi liquid description of the interacting Fermi gas usually works in 2D or 3D. However, it breaks down in 1D and the systems become Luttinger liquids.
Fortunately, there have been a number of many-body techniques suitable to study various 1D problems\cite{Giamarchi}, which make the generalization of the MZMs in 1D models accessible.
On the other hand, the interaction will modify topological systems violently, e.g. the non-interacting classification of fermionic systems\cite{Ryu2008,Ryu2010,Kitaev2009} will ``collapse''
and there exists a continuous path connecting trivial and topological phases in 1D\cite{Fidkowski10}.


Kitaev chain\cite{Kitaev} is a prototype of 1D systems possessing MZMs at the two edges.
The non-interacting Kitaev model was initially solved in a ring with periodic boundary condition. The edge state was then proposed to exhibit MZM. The model has been generalized to interacting case with nearest neighboring repulsive interaction.
The interacting Kitaev model does not have analytic solutions in general cases except for a set of specially tuned parameters\cite{Katsura,Rahmani2}. The model can also be studied by numerical methods\cite{Rahmani2,Thomale2013,Gergs}.
In general, interacting effects on MZMs have been investigated in various systems, e.g. nanowires\cite{Gangadharaiah,Stoudenmire,Thomale,Manolescu,Chan}, multiband nanowires\cite{Lutchyn}, helical liquids\cite{Sela},
two-leg ladders\cite{Cheng}, Josephson junctions\cite{Hassler}, Abrikosov vortex lattice\cite{Chiu} and topological insulator/superconductor heterostructure\cite{Hung}. The interplay of disorder and interaction has  also been analyzed\cite{Lobos,Crepin}.
The MZM is stable against weak perturbations including the interaction and disorder.
However, the generic interaction effect remains an open question, although lots of efforts have been made,
which includes the exact solution\cite{Mazza},topological classification\cite{Fidkowski10,Fidkowski11}, entanglement entropy investigation\cite{Turner}, many-body MZM operator\cite{Goldstein,Kells},
super-symmetry approaches\cite{Grover,Ulrich,Rahmani1,Rahmani2} and parafermion edge zero mode\cite{Fendley,Clarke,Klinovaja,Jermyn,Alexandradinata}.

In this paper, we shall first study non-interacting Kitaev chain of length $L$ with open boundary condition by using an analytic method, which is accessible at zero chemical potential or at a symmetric point of the pairing and the hopping amplitudes, $\Delta = t$.
We propose a correlation function of the two Majorana operators as a long range order parameter to describe non-trivial topological state with edge MZMs and calculate the long range correlation function explicitly. We then study Kitaev model with nearst neighboring repulsion interaction in open boundary condition by using density matrix renormalization group (DMRG) method.  We show that the qualitative feature of the long range correlation remain unchanged in the interacting systems provided that the system is in the topological non-trivial phase.  The phase diagram in the interacting model will also be discussed. This paper is complementary to our previous work about exact solution to interacting Kitaev chain at symmetric point\cite{Miao}. In this paper we show how to diagonalize the non-interacting Hamiltonian in details and extend the region of phase diagram away from $\mu=0$.

This paper is organized as follows. In Section~\ref{model}, the model Hamiltonian are presented and Majorana fermion representation is introduced.
In Section~\ref{noninteracting}, we study non-interacting models by using analytic solutions. A single-particle correlation function is introduced and its edge component is used to describe the topological order.
In Section~\ref{Sec:DMRG}, numerical DMRG analysis is carried out to study interacting systems.
Section~\ref{conclusion} is devoted to discussions.

\section{Model}\label{model}

Without loss of generality, we consider a chain of spinless fermions with open boundary condition. The Hamiltonian of such an interacting Kitaev chain is
\begin{eqnarray}
H&=&\sum_{j=1}^{L-1}\left[-t\left(c_{j}^{\dagger}c_{j+1}+h.c.\right)+U\left(2n_{j}-1\right)\left(2n_{j+1}-1\right)\right.\nonumber\\
 &&\left.-\Delta\left(c_{j}^{\dagger}c_{j+1}^{\dagger}+h.c.\right) \right] -\mu \sum_{j=1}^{L}\left(n_{j}-{1\over2}\right),\label{eq:H}
\end{eqnarray}
where $c_{j}(c_{j}^{\dagger})$ is fermion annihilation (creation) operator on site $j$, $n_{j}=c_{j}^{\dagger}c_{j}$ is the fermion number operator,
$t$ is the hopping matrix element, and $\Delta$ is the $p$-wave superconducting pairing potential induced by the proximity effect, $\mu$ is the chemical potential controlling the electron density, and $U$ is the nearest neighbor interaction.
One can always choose $\Delta$ real and non-negative by the global transformation $c_{j}\rightarrow e^{i\varphi}c_{j}$.
Similarly, one can study the case of $t\geq0$ and $\mu\geq0$ only, since the parameter transformations $t\to -t$ and $\mu\to -\mu$ can be realized by by the gauge transformation $c_{j}\rightarrow i\left(-1\right)^{j}c_{j}$
and particle-hole cojugation $c_{j}\rightarrow\left(-1\right)^{j}c_{j}^{\dagger}$ respectively. Note that all these transformations will keep other parameters unchanged.
In this paper, we only consider repulsive nearest neighbor interaction with $U\geq0$. When $U=0$, this model will reduce to the usual (non-interacting) Kitaev chain\cite{Kitaev}.

The Hamiltonian has the fermion number parity $Z_{2}^{f}$ symmetry, which is defined as
\begin{equation}\label{def:Z2f}
Z_{2}^{f}=e^{i\pi\sum_{j}n_{j}}=\left(-1\right)^{\hat{N}},
\end{equation}
where $\hat{N}=\sum_{j}n_{j}$ is the total fermion number, and it is obvious that $(Z_{2}^f)^2=1$ and $[H,Z_{2}^f]=0$. $Z_{2}^{f}$ conserves in the whole parameter space. In the presence of the pairing potential $\Delta$, the total fermion number is not conserved but only conserved modulo $2$.

\subsection{Majorana fermion representation}\label{majorana}

We shall use the Majorana fermion representation to investigate the interacting Kitaev chain. Following Katsura {\it et al.}\cite{Katsura}, we split one complex fermion operator into two Majorana fermion operators
\numparts
\begin{eqnarray}
c_{j} & = & \frac{1}{2}\left(\lambda_{j}^{1}+i\lambda_{j}^{2}\right),\\
c_{j}^{\dagger} & = & \frac{1}{2}\left(\lambda_{j}^{1}-i\lambda_{j}^{2}\right).
\end{eqnarray}
\endnumparts
The Majorana fermion operators are real
\begin{equation}
\left(\lambda_{j}^{a}\right)^{\dagger}=\lambda_{j}^{a},
\end{equation}
and satisfy the anticommutation relations
\begin{equation}
\left\{ \lambda_{j}^{a},\lambda_{l}^{b}\right\} =2\delta_{ab}\delta_{jl},
\end{equation}
where $a,b=1,2$. In the Majorana fermion representation, the Hamiltonian
of the interacting Kitaev chain becomes
\begin{eqnarray}
H&=&\sum_{j=1}^{L-1} [-\frac{i}{2}\left(t+\Delta\right)\lambda_{j+1}^{1}\lambda_{j}^{2}-\frac{i}{2}\left(t-\Delta\right)\lambda_{j}^{1}\lambda_{j+1}^{2} \nonumber\\
&&-U\lambda_{j}^{1}\lambda_{j}^{2}\lambda_{j+1}^{1}\lambda_{j+1}^{2}] -\frac{i}{2}\mu \sum_{j=1}^{L}\lambda_{j}^{1}\lambda_{j}^{2}.
\end{eqnarray}

\section{Non-interacting Kitaev chains}\label{noninteracting}
 In this section, we consider the non-interacting Kitaev chains with open boundary condition and discuss the relations among the topological degeneracy, the Majorana zero mode, and the edge correlation functions.
We shall use analytic method to exactly solve the two non-interacting cases with $\Delta=t,\, U=0$ and $\mu=0,\, U=0$ by the singular value decomposition (SVD) in Majorana fermion representation.

\subsection{Non-interacting chains with $\Delta=t$}
In this case, the transition between the topological superconductor and the trivial superconductor can be studied by tuning the chemical potential $\mu$. The non-interacting Hamiltonian $H_{\mu}$ is quadratic in $\lambda_{j}^{1}$ and $\lambda_{j}^{2}$ and is given by
\begin{eqnarray}
H_{\mu} & = & \frac{i}{2}\left[\sum_{j=1}^{L-1}-2t\lambda_{j+1}^{1}\lambda_{j}^{2}-\sum_{j=1}^{L}\mu\lambda_{j}^{1}\lambda_{j}^{2}\right]\nonumber \\
 & = & \frac{i}{2}\sum_{j,l=1}^{L}\lambda_{j}^{1}B_{jl}\lambda_{l}^{2},
\end{eqnarray}
where $B$ is a $L\times L$ real matrix,
\begin{equation}\label{eq:matrixB}
B=-\left(\begin{array}{ccccc}
\mu & 0\\
2t & \mu & 0\\
 & \ddots & \ddots & \ddots\\
 &  & 2t & \mu & 0\\
 &  &  & 2t & \mu
\end{array}\right).
\end{equation}
With the help of SVD, $B=U\Lambda V^T$, where $\Lambda$ is a real diagonal matrix, $U$ and $V$ are real orthogonal matrices, $H_{\mu}$ can be diagonalized as follows,
\begin{eqnarray}
H_{\mu} & = & \frac{i}{2}\sum_{k}\lambda_{k}^{1}\Lambda_{k}\lambda_{k}^{2}\nonumber \\
 & = & \sum_{k}\Lambda_{k}\left(c_{k}^{\dagger}c_{k}-\frac{1}{2}\right),
\end{eqnarray}
where $\Lambda_{k}\geq 0$ are singular values of the matrix $B$, $c_{k}  =  \frac{1}{2}\left(\lambda_{k}^{1}+i\lambda_{k}^{2}\right)$ and $c_{k}^{\dagger}  = \frac{1}{2}\left(\lambda_{k}^{1}-i\lambda_{k}^{2}\right)$
are the complex fermion operators.

In the weak pairing region, $\mu<2t$, we find that (See Appendix \ref{ED:non-interacting} for details) the smallest singular value $\Lambda_{k}$ is nonzero given by
\begin{equation}\label{eq:Lambdak0}
\Lambda_{k_{0}}=\Bigl(\frac{2t}{\mu}-\frac{\mu}{2t}\Bigr)\Bigl(\frac{\mu}{2t}\Bigr)^{L},
\end{equation}
and the corresponding matrix elements
\numparts\label{eq:UVk0}
\begin{eqnarray}
U_{jk_{0}} & = & A_{k_{0}}\sinh v\left(L+1-j\right),\\
V_{jk_{0}} & = & A_{k_{0}}\sinh v j,
\end{eqnarray}
\endnumparts
where $A_{k_{0}}=2e^{-v L}\left(1-e^{-2v}\right)^{1/2}$ is the normalization factor,
and $v$ is a positive real number determined by Eq.~\eref{eq:vA}.

It is worth noting that a similar model has been solved by Katsura {\it et al.}\cite{Katsura} using SVD. In their case, the chemical potential is half of the bulk's value at edge, $\mu_{1}=\mu_{L}=\mu/2$, resulting in $\Lambda_{k_{0}}=0$.

\subsubsection{Topological degeneracy and the edge mode}\label{Majorana}

It is well known that there exist two topologically distinct phases in the non-interacting Kitaev chain model\cite{Kitaev,Qi,Bernevig}.
For strong pairing $\mu>2t$, the system is in the trivial superconducting state, while for weak pairing $\mu<2t$, the system is in the topological superconducting state.

In the trivial superconducting state, the energy spectrum is gapped and the ground state is non-degenerate.
However, in the topological superconductor, the energy gap between the ground state $\left|0\right\rangle $ and the first excited state $\left|1\right\rangle \equiv c_{k_{0}}^{\dagger}\left|0\right\rangle$ is $\Lambda_{k0}$ given in Eq.~\eref{eq:Lambdak0},
 approaches to zero with the exponential factor $e^{-L\ln(2t/\mu)}$ in the large $L$ limit.
Thus, the $k_0$-mode is a {\em zero mode} and the topological superconductor has two-fold degenerate ground states in thermodynamic limit. In other words, it is a gapped system with two-fold topological degeneracy.

\begin{figure}[hptb]
\begin{center}
\includegraphics[width=7.2cm]{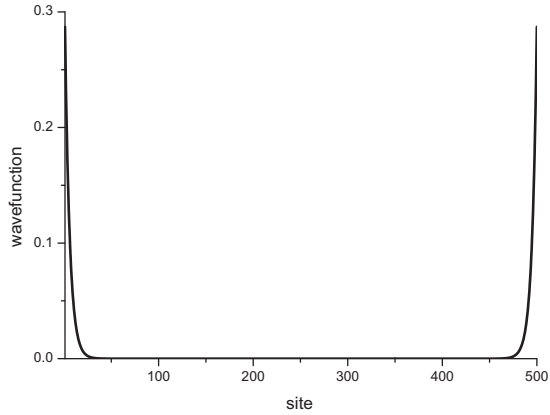}
\end{center}
\caption{
The particle wavefunction $\left\langle 0\right|c_{j}c_{k_{0}}^{\dagger}\left|0\right\rangle$ for the $k_0$-mode with $L=500$ and $v=0.2$. }
\label{fig:wavefunction}
\end{figure}

Now we shall check that the first excited state $\left|1\right\rangle$ is an edge mode. It is a single particle (hole) excited state. The particle and hole parts of the wavefunction read
\numparts
\begin{eqnarray}
\left\langle 0\right|c_{j}\left|1\right\rangle & = &\left\langle 0\right|c_{j}c_{k_{0}}^{\dagger}\left|0\right\rangle  = \frac{1}{2}\left(U_{jk_{0}}+V_{jk_{0}}\right)\nonumber \\
 & = & \frac{A_{k_{0}}}{2}\left[\sinh v\left(L+1-j\right)+\sinh v j\right]
\end{eqnarray}
and
\begin{eqnarray}
\left\langle 0\right|c_{j}^{\dagger}\left|1\right\rangle & = &\left\langle 0\right|c_{j}^{\dagger}c_{k_{0}}^{\dagger}\left|0\right\rangle  = \frac{1}{2}\left(U_{jk_{0}}-V_{jk_{0}}\right)\nonumber \\
 & = & \frac{A_{k_{0}}}{2}\left[\sinh v\left(L+1-j\right)-\sinh v j\right]
\end{eqnarray}
\endnumparts
respectively, where Eq.~\eref{eq:UVk0} and Eq.~\eref{eq:UVA1} have been used in the derivation.
It is easy to see that this zero mode has a complex wave vector $k_{0}=\pi+iv$ and the wavefunction is well localized at edges with localization length $v^{-1}$ as demonstrated in Fig.~\ref{fig:wavefunction}.

Now we would like to examine that the $k_0$ mode is indeed a Majorana mode, say, $c_{k_{0}}^{\dagger}=\pm c_{k_{0}}$, namely, it coincides to its antiparticle.
Using Eq.~\eref{eq:UVA1}, we have
\begin{equation}
c_{k_{0}}^{\dagger}=\frac{1}{2}\left(\lambda_{k_{0}}^{1}-i\lambda_{k_{0}}^{2}\right)=\frac{1}{2}\sum_{j=1}^{L}\left(U_{jk_0}\lambda_{j}^{1}-iV_{jk_0}\lambda_{j}^{2}\right).
\end{equation}
By Eq.~\eref{eq:UVk0}, we find that
\begin{equation}
c_{k_{0}}^{\dagger}=\left\{\begin{array}{cc} c_{k_{0}}, & j\ll v^{-1},\\ -c_{k_{0}}, & L+1-j\ll v^{-1}. \end{array} \right.
\end{equation}
So that there exists one Majorana mode with $c_{k_{0}}^{\dagger}=c_{k_{0}}$ at the edge $j=1$ and another Majorana mode with $c_{k_{0}}^{\dagger}=-c_{k_{0}}$ at the edge $j=L$.

\subsubsection{Fermion number parity and edge correlation function}\label{Sec:FP}

There are two characterizing features for topological ordered systems, (base-manifold dependent) ground state degeneracy and gapless edge states.


We note the ground state $\left|0\right\rangle $ and the excited state $\left|1\right\rangle $ have opposite fermion number parity
\begin{equation}
\left\langle 1\right| Z_{2}^{f}\left|1\right\rangle = \left\langle 0\right| c_{k_{0}} Z_{2}^{f}c_{k_{0}}^{\dagger}\left|0\right\rangle = -\left\langle 0\right|Z_{2}^{f}\left|0\right\rangle .
\end{equation}
In the thermodynamic limit, the first excited $\left|1\right\rangle $ is degenerate with the ground state $\left|0\right\rangle $.

We define the following single-particle correlation function at two sites $j$ and $l$,
\numparts\label{eq:defG1L}
\begin{equation}
G_{jl}=\left\langle i\lambda_{j}^{1}\lambda_{l}^{2}\right\rangle ,
\end{equation}
where the imaginary $i$ is introduced to make $G_{jl}$ Hermitian. Especially, the edge component of $G_{jl}$ is given when $j=1$ and $l=L$,
\begin{equation}
G_{1L}=\left\langle i\lambda_{1}^{1}\lambda_{L}^{2}\right\rangle.
\end{equation}
\endnumparts
Note that the correlation function $G_{jl}$ is a block of single-particle(hole) density of matrix, which can be generalized to interacting systems and reflects the site-distribution of single-particle component in a many-particle wavefunction.
As long as the bulk is uniform, the finite value of $G_{1L}$ in the thermodynamic limit reflects the existence of edge modes.

The edge correlation function $G_{1L}$ is easy to calculate in the case of $\Delta=t$ and $U=0$, and is given for the ground state $\left|0\right\rangle$ by
\begin{equation}
G_{1L}=\left\langle 0\right|i\lambda_{1}^{1}\lambda_{L}^{2}\left|0\right\rangle=-\sum_{k}U_{1k}V_{Lk}.
\end{equation}
When $\mu\geq2t$,
\begin{equation}
G_{1L}=\left\langle 0\right|i\lambda_{1}^{1}\lambda_{L}^{2}\left|0\right\rangle=-\sum_{k}A_{k}^{2}\delta_{k}\sin^{2}kL.
\end{equation}
As proved by Lieb et al.\cite{Lieb} , this summation is of order of $O\left(1/L\right)$. When $\mu<2t$,
\begin{eqnarray}
G_{1L} & = & \left\langle 0\right|i\lambda_{1}^{1}\lambda_{L}^{2}\left|0\right\rangle = -U_{1k_{0}}V_{Lk_{0}}-\sum_{k}U_{1k}V_{Lk}\nonumber \\
 & = & -A_{k_{0}}^{2}\sinh^{2} v L-\sum_{k}A_{k}^{2}\delta_{k}\sin^{2}kL\nonumber \\
 & = & -\left[1-\left(\frac{\mu}{2t}\right)^{2}\right]+O\left(1/L\right).
\end{eqnarray}
The nonvanishing value of $G_{1L}$ for $\mu<2t$ in the thermodynamic limit reflects the topological order in the topological superconductor state.
In this topological phase, we can also calculate edge correlation function $G_{1L}$ for the topological degenerate state $\left|1\right\rangle $.
\begin{eqnarray}
G_{1L} & = & \left\langle 1\right|i\lambda_{1}^{1}\lambda_{L}^{2}\left|1\right\rangle  =  U_{1k_{0}}V_{Lk_{0}}-\sum_{k}U_{1k}V_{Lk}\nonumber \\
 & = & A_{k_{0}}^{2}\sinh^{2} v L-\sum_{k}A_{k}^{2}\delta_{k}\sin^{2}kL\nonumber \\
 & = & \left[1-\left(\frac{\mu}{2t}\right)^{2}\right]+O\left(1/L\right).
\end{eqnarray}
Thus, for a generic ground state $\left|GS\right\rangle$, the edge correlation function in the thermodynamic limit is given by
\begin{equation}
\lim_{L\to\infty}G_{1L}\propto \left\{ \begin{array}{cc} 1-\left(\frac{\mu}{2t}\right)^{2}, & \mu<2t, \\  0, & \mu\geq 2t.\end{array} \right.
\end{equation}
Note that the nonzero contribution $U_{1k_{0}}V_{Lk_{0}}$ comes from the Majorana zero mode $k_{0}$. Other modes mainly distribute in the bulk and the contributions to $G_{1L}$ is of order of $O\left(1/L\right)$, which is neglectable in the thermodynamic limit.
At the quantum critical point $\mu=2t$, we have $v=0$ and the wave vector of the Majorana zero mode becomes real $k_{0}=\pi$. The $k_0$-mode is no longer localized at edges but merges into the bulk, resulting in vanishing edge correlation function $G_{1L}$.
In the quantum critical region,
\begin{equation}
G_{1L}\propto (2t-\mu)^z,
\end{equation}
with critical exponent $z=1$.

\begin{figure}[hptb]
\begin{center}
\includegraphics[width=8.4cm]{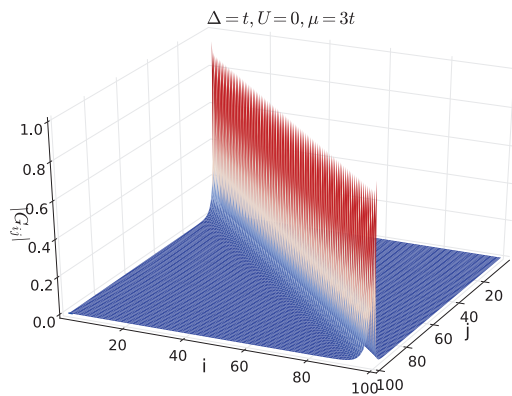}
\end{center}
\caption{
Correlation function $|G_{ij}|$ for a topologically trivial state, $\Delta=t,\mu=3t,U=0$. }
\label{fig:GijSC1}
\end{figure}

\begin{figure}[hptb]
\begin{center}
\includegraphics[width=8.4cm]{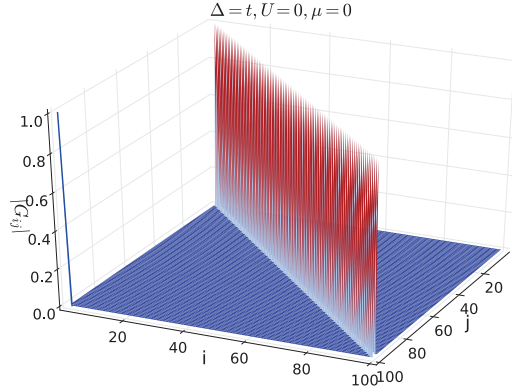}
\end{center}
\caption{
Correlation function $|G_{ij}|$ for a topologically nontrivial state, $\Delta=t,\mu=0,U=0$.  }
\label{fig:GijTSC1}
\end{figure}

Now we would like to examine the behavior of $G_{ij}$ inside the bulk, which can be done numerically. Two topologically distinct examples are investigated and shown in Fig.~\ref{fig:GijSC1} and Fig.~\ref{fig:GijTSC1} respectively.
The first example is given by $\Delta=t,\mu=3t,U=0$, which is in the topologically trivial phase, where a peak appears at short range with $i\sim j$ while long range correlation is absent.
The second example is given by $\Delta=t,\mu=t,U=0$, which is in the nontrivial topological superconductor phase. There exhibits a long range peak at $i=1$ and $j=L$, and long range correlation is still absent inside the bulk.
We note the edge correlation is not symmetric or antisymmetric, i.e. $G_{1L}\neq \pm G_{L1}$. Hence there is no peak at $i=L$ and $j=1$. If we use parameters with $t<0$, the peak will appear at $i=L$ and $j=1$. So it is a matter of choice. The point is there is a edge correlation function corresponding to the Majorana zero mode.

Therefore, we propose to use the edge correlation function $G_{1L}$ to characterize the topological order and emerged edge states. We shall examine this for the  non-interacting systems with different parameters in the next subsection and for the interacting systems in the next section.

\subsection{Non-interacting chains with $\mu=0$}\label{Sec:HDelta}
In this subsection, we utilize non-interacting Kitaev chains with $\mu=0$ to study how topological order will vanish as the superconducting gap $\Delta$ approaches zero.
The Hamiltonian now reads
\begin{equation}
H_{\Delta}=\frac{i}{2}\sum_{j=1}^{L-1}\left[-\left(t+\Delta\right)\lambda_{j+1}^{1}\lambda_{j}^{2}-\left(t-\Delta\right)\lambda_{j}^{1}\lambda_{j+1}^{2}\right].
\end{equation}
We are able to diagonalize the Hamiltonian $H_{\Delta}$ by SVD as before. There exist two kinds of modes in this situation. For the first kind of modes, the two orthogonal matrices $U$ and $V$ are found to be
\numparts\label{eq:UVkI}
\begin{eqnarray}
U_{jk^{I}} & = & \cases{
0, & j = odd,\\
A_{k^{I}}\sin k^{I}j, & j = even,}\\
V_{jk^{I}} & = & \cases{
-A_{k^{I}}\delta_{k^{I}}\sin k^{I}\left(L+1-j\right), & j=odd,\\
0, & j = even.}
\end{eqnarray}
\endnumparts
The second kind of modes is given by
\numparts\label{eq:UVkII}
\begin{eqnarray}
U_{jk^{II}} & = & \cases{
A_{k^{II}}\sin k^{II}\left(L+1-j\right), & j=odd,\\
0, & j=even,
}\\
V_{jk^{II}} & = & \cases{
0, & j=odd,\\
-A_{k^{II}}\delta_{k^{II}}\sin k^{II}j, & j=even.
}
\end{eqnarray}
\endnumparts
Here the normalization factors are given by
\begin{equation}
A_{k}=2\left[L+1-\frac{\sin2k\left(L+1\right)}{\sin2k}\right]^{-1/2},
\end{equation}
and
\begin{equation}
\delta_{k}=sgn \left[ \frac{\cos k}{\cos k\left(L+1\right)} \right].
\end{equation}
Corresponding singular values are given by
\begin{equation}
\Lambda_{k}=\sqrt{\left(2t\cos k\right)^{2}+\left(2\Delta\sin k\right)^{2}}.
\end{equation}
The wave vector $k^I$'s are given by the following equation,
\begin{equation}
\frac{\sin k^{I}\left(L+2\right)}{\sin k^{I}L}=-\frac{t-\Delta}{t+\Delta},
\end{equation}
and $k^{II}$'s are determined by
\begin{equation}
\frac{\sin k^{II}\left(L+2\right)}{\sin k^{II}L}=-\frac{t+\Delta}{t-\Delta}.
\end{equation}
Besides $L-1$ real $k^{II}$'s, there exists a single complex $k^{II}$ in the second kind modes,
\begin{equation}
k_{0}^{II}=\frac{\pi}{2}+iv,
\end{equation}
with $v$ determined by
\begin{equation}
\frac{\sinh v\left(L+2\right)}{\sinh vL}=\frac{t+\Delta}{t-\Delta}.
\end{equation}
For this $k_0^{II}$ mode we have
\numparts\label{eq:UVk0II}
\begin{eqnarray}
U_{jk_{0}^{II}} & = & \cases{
A_{k_{0}^{II}}\left(-1\right)^{\frac{L+1-j}{2}}\sinh v\left(L+1-j\right) & j=odd,\\
0 & j=even,
}\\
V_{jk_{0}^{II}} & = & \cases{
0 & j=odd,\\
-A_{k_{0}^{II}}\left(-1\right)^{-\frac{L-j}{2}}\sinh vj & j=even.
}
\end{eqnarray}
\endnumparts
Then the normalization factor can be written explicitly,
\begin{equation}
A_{k_{0}^{II}}=2e^{- v L}\left(1-e^{-4v}\right)^{1/2},
\end{equation}
and the singular value reads
\begin{equation}
\Lambda_{k_{0}^{II}}=\frac{2\Delta}{t+\Delta}\left(\frac{t-\Delta}{t+\Delta}\right)^{L/2}.
\end{equation}
It is easy to see that the singular value of $k_{0}^{II}$ mode vanishes in the thermodynamic limit,
\begin{equation}
\lim_{L\rightarrow\infty}\Lambda_{k_{0}^{II}}=0.
\end{equation}
The (single particle) wavefunction of this zero mode is given by
\begin{eqnarray}
 &  & \left\langle 0\right|c_{j}c_{k_{0}^{II}}^{\dagger}\left|0\right\rangle  =  \frac{1}{2}\left(U_{jk_{0}^{II}}+V_{jk_{0}^{II}}\right)\nonumber \\
 & = & \frac{A_{k_{0}^{II}}}{2}\cases{
\left(-1\right)^{\frac{L+1-j}{2}}\sinh v\left(L+1-j\right), & j=odd,\\
-\left(-1\right)^{-\frac{L-j}{2}}\sinh vj, & j=even,
}
\end{eqnarray}
which has nonzero value only near the edge in the thermodynamic limit. Similarly, one can verify that $c_{k_{0}^{II}}^{\dagger}=\pm c_{k_{0}^{II}}$ at edges. Hence the $k_{0}^{II}$-mode is the Majorana zero mode localized at edges.
When $\Delta\to 0$, the wave vector of the zero mode becomes real $k_{0}^{II} =\frac{\pi}{2}$ and the Majorana zero mode is no longer localized at edges.
This is consistent with the condition for the boundary Majorana fermion argued by Kitaev\cite{Kitaev}, i.e. the presence of an arbitrary small superconducting gap $\Delta$.

Now we compute the edge correlation function $G_{1L}$ for the ground state $\left|0\right\rangle$,
\begin{eqnarray}
G_{1L} & = & \left\langle 0\right| i\lambda_{1}^{1}\lambda_{L}^{2} \left|0\right\rangle = - U_{1k_{0}^{II}}V_{Lk_{0}^{II}}-\sum_{k}U_{1k}V_{Lk}\nonumber \\
 & = & \left(-1\right)^{L/2}A_{k_{0}^{II}}^{2}\sinh^{2} v L+\sum_{k}A_{k}^{2}\delta_{k}\sin^{2}kL\nonumber \\
 & = & \left(-1\right)^{L/2}\left[1-\left(\frac{t-\Delta}{t+\Delta}\right)^{2}\right]+O\left(1/L\right),
\end{eqnarray}
and for the topological degenerate state $\left|1\right\rangle= c_{k_{0}^{II}}^{\dagger}\left|0\right\rangle$,
\begin{eqnarray}
G_{1L} & = & \left\langle 1\right|i\lambda_{1}^{1}\lambda_{L}^{2}\left|1\right\rangle  =  U_{1k_{0}^{II}}V_{Lk_{0}^{II}}-\sum_{k}U_{1k}V_{Lk}\nonumber \\
 & = & -\left(-1\right)^{L/2}A_{k_{0}}^{2}\sinh^{2} v L+\sum_{k}A_{k}^{2}\delta_{k}\sin^{2}kL\nonumber \\
 & = & -\left(-1\right)^{L/2}\left[1-\left(\frac{t-\Delta}{t+\Delta}\right)^{2}\right]+O\left(1/L\right).
\end{eqnarray}
For small but finite $\Delta$, we have
\begin{equation}
G_{1L}\propto \Delta^z,
\end{equation}
with critical exponent $z=1$. Thus the edge correlation function vanished as $\Delta\to 0$.

\section{Interacting Kitaev chains: DMRG analysis}\label{Sec:DMRG}
In this section, we shall study interacting Kitaev chains by carrying out DMRG calculations in the language of matrix product states\cite{Schollwck}
with various model parameters in Hamiltonian \eref{eq:H} and system size up to $L=140$.
We compute the energy of low lying states, local particle density, as well as the single-particle correlation function $G_{ij}$.

\begin{figure}[hptb]
\begin{center}
\includegraphics[width=14cm]{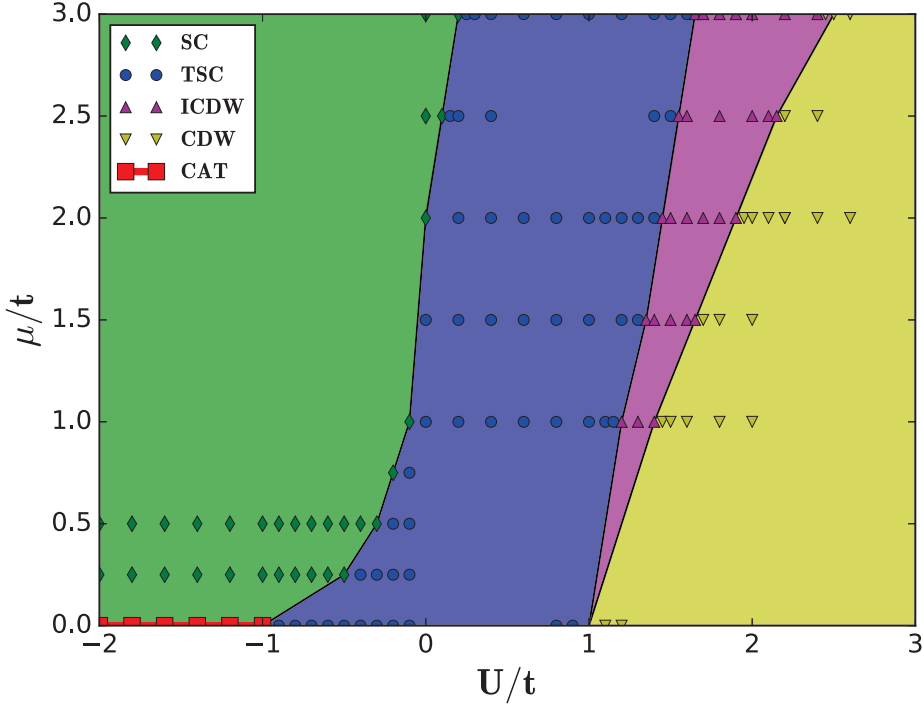}
\end{center}
\caption{
Phase diagram for the interacting Kitaev chain with $\Delta=t$. SC stands for trivial superconductor, TSC stands for topological superconductor, CDW stands for charge density wave, ICDW stands for incommensurate charge density wave and CAT stands for Shr\"{o}dinger-cat-like (CAT) state.
Data points are obtained within DMRG for different system sizes. Rhombuses denote SC states, circles denote TSC states, up-triangles denote ICDW states, down-triangles denote CDW states and squares denote CAT states.
}
\label{fig:phase}
\end{figure}

{\em Phase diagrams.} Fig.~\ref{fig:phase} displays the phase diagram at $\Delta=t$ obtained from the combination of exact solutions and DMRG calculations.
As a function of $\mu$ and $U$, there are five distinct phases, trivial superconductor (SC), topological superconductor (TSC), commensurate charge density wave (CDW), incommensurate charge density wave (ICDW) and Shr\"{o}dinger-cat-like state (CAT).
The five different phases are separated from each other by critical lines. Such a phase diagram is consistent with previous studies\cite{Katsura,Rahmani2,Thomale2013} except the CAT states at $\mu=0$ obtained by exact solution\cite{Miao}

The TSC phase is detected by the two-fold degenerate ground states with opposite fermion number parity $Z_{2}^{f}$ and CAT phase is the two-fold degenerate ground states with opposite particle-hole symmetry $Z_{2}^{p}$. In contrast, the two ground states of CDW and ICDW phase have the same $Z_{2}^{f}$.
In practice, we compute the matrix elements for $Z_{2}^f$ or $Z_{2}^{p}$ in the subspace spanned by the two lowest lying states, $\left|0\right\rangle$ and $\left|1\right\rangle$, and diagonalize the $2\times 2$ matrix to obtain two eigenvalues.
The distinction between ICDW and CDW can be made through local particle density and its Fourier transformation. For a CDW state, there exists a single peak at $Q=\pi$, while for a ICDW state, there appear two peaks in the Fourier spectrum.

When $\mu=0$, as $U$ increases, the ground state changes from CAT to TSC and to CDW directly via the critical point $U=\pm t$. When $\mu>0$, as $U$ increase, the ground state changes from SC to TSC, ICDW and to CDW in the large $U$ limit.

\begin{figure}[hptb]
\begin{center}
\includegraphics[width=8.4cm]{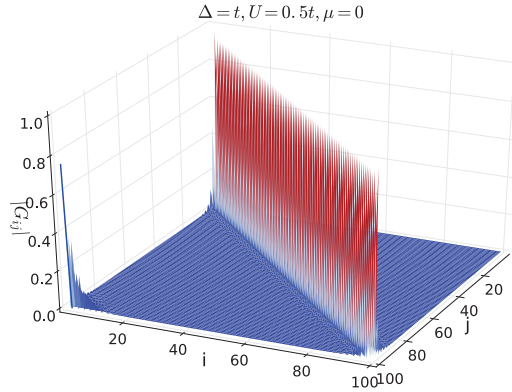}
\end{center}
\caption{
Single-particle correlation function $G_{ij}$ for the TSC ground state with $\Delta=t$, $\mu=0$ and $U=0.5t$. The system size is $L=100$.
}
\label{fig:GijU0.5mu0}
\end{figure}
\begin{figure}[hptb]
\begin{center}
\includegraphics[width=8.4cm]{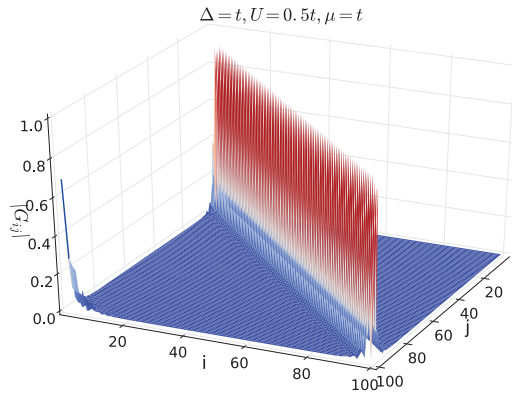}
\end{center}
\caption{
Single-particle correlation function $G_{ij}$ for the TSC ground state with $\Delta=t$, $\mu=t$ and $U=0.5t$. The system size is $L=100$.
}
\label{fig:GijU0.5mu1}
\end{figure}

{\em Single-particle correlation function $G_{ij}$.} We also compute the single-particle correlation function $G_{ij}$ defined in Eq.~\eref{eq:defG1L} for ground states.
Similar to exactly solvable systems shown in Fig.~\ref{fig:GijSC1} and Fig.~\ref{fig:GijTSC1}, long range correlation is absent inside the bulk. When the system is in the TSC phase,
there exists a single long range peak at $i=1$ and $j=L$. Fig.~\ref{fig:GijU0.5mu0} and Fig.~\ref{fig:GijU0.5mu1} demonstrate two TSC states with $\Delta=t,\mu=0,U=0.5t$ and $\Delta=t,\mu=t,U=0.5t$ respectively.
So that $G_{ij}$ serves an efficient measurement for edge states and thereby the topological order.

\begin{figure}[hptb]
\begin{center}
\includegraphics[width=8.4cm]{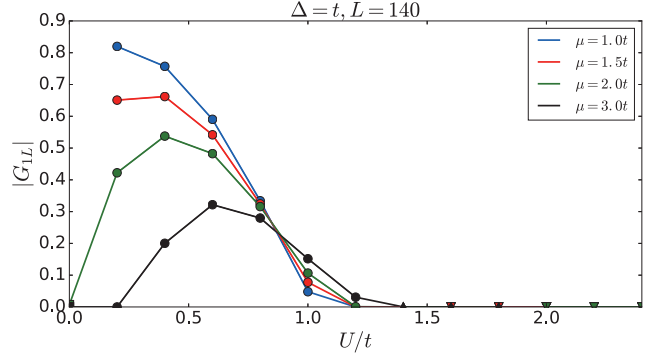}
\end{center}
\caption{
Ground state edge correlation function $G_{1L}$ as function of $\mu$ and $U$. $\Delta=t$ and the system size is $L=140$. Squares denote SC states, circles denote TSC states, up-triagnles denote ICDW states, and down-triangles denote CDW states.
}
\label{fig:G1L140MPS}
\end{figure}

{\em Edge correlation function $G_{1L}$.} The nonvanishing edge correlation function $G_{1L}$ characterizes the topological order.
We fix $\Delta=t$ and study $G_{1L}$ as a function of $\mu$ and $U$. The result is plotted in Fig.~\ref{fig:G1L140MPS}. The value of $G_{1L}$ is finite in TSC phase and vanishes in other topologically trivial phases. Thus this order parameter is valid both in the non-interacting and interacting systems to study the topological order.

{\em Local density of states.} We can distinguish the ICDW and CDW phases by observing their local density distribution and corresponding Fourier spectrum.  When the ground state is a CDW, its Fourier spectrum will have a single peak at $Q=\pi$; while for a ICDW state there are two peaks.
\par
\begin{figure}[hptb]
\begin{center}
\includegraphics[width=8.4cm]{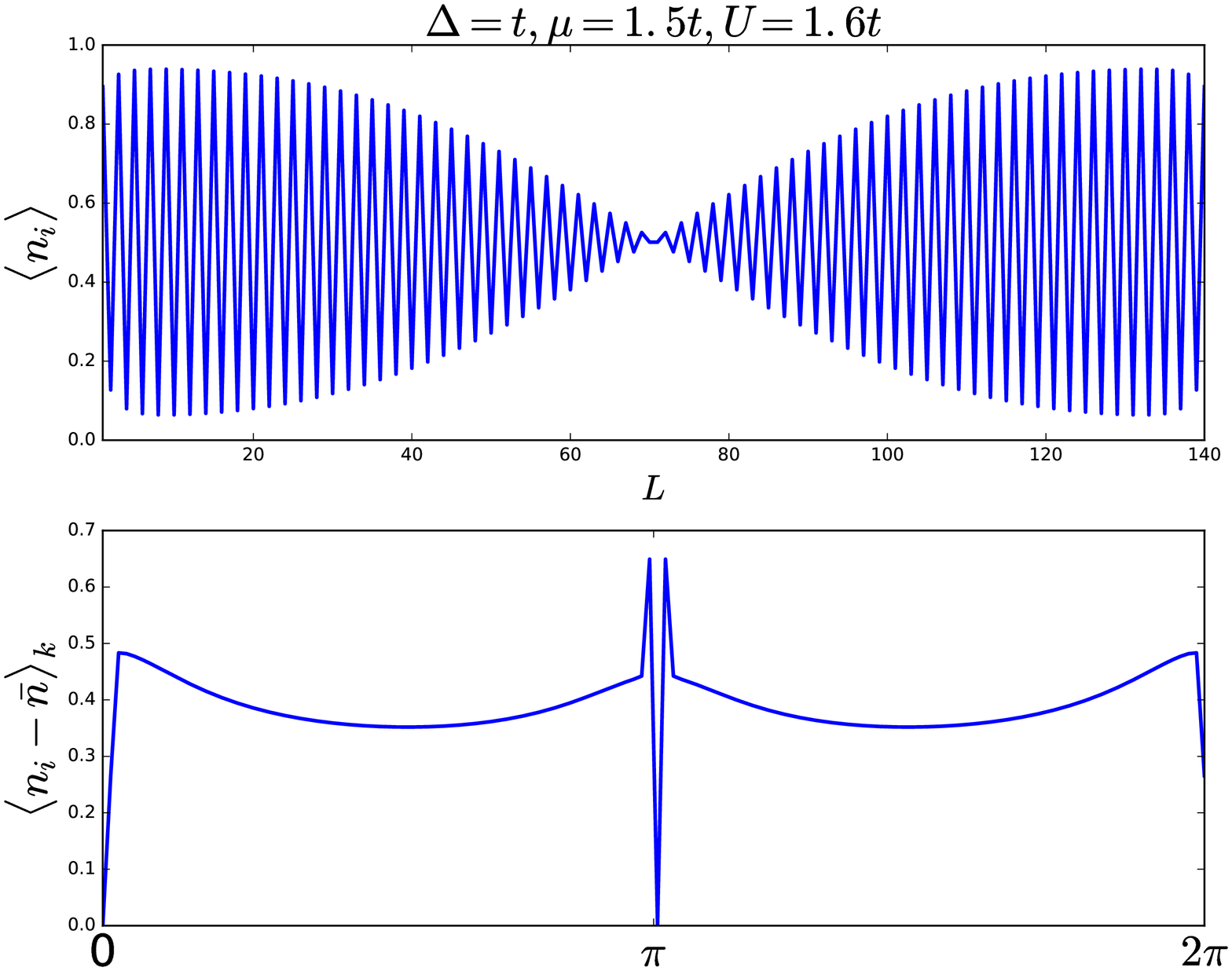}
\includegraphics[width=8.4cm]{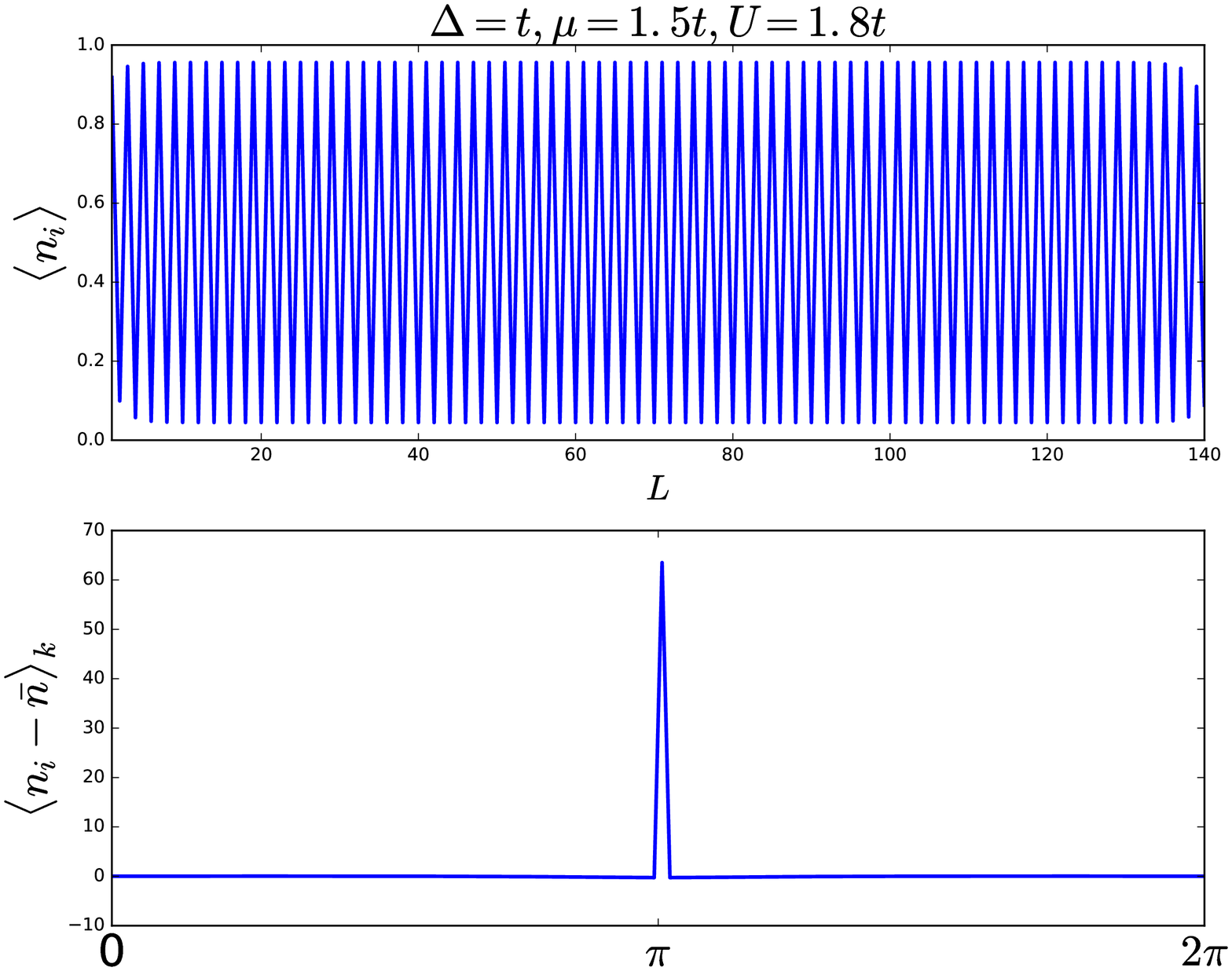}
\end{center}
\caption{Local density distribution and density spectrum. In up figure the local density of ICDW oscillates nonuniformly and its Fourier spectrum has two peaks near $Q=\pi$. In bottom figure the local density of CDW forms a bipartite lattice and the Fourier spectrum has single peaks at $Q=\pi$.}
\label{fig:CDW}
\end{figure}

{}
For various model parameters, we use the DMRG method to obtain the ground state $|0\rangle$ and local density $\langle{}0|\hat{n}_{j}|0{}\rangle$ for each site $j$. The Fourier spectrum is obtained by taking fast Fourier transformation of the local density distribution, whose average value has been subtracted.  Here we show two typical figures of ICDW and CDW in Fig.~\ref{fig:CDW}.

\section{Conclusion}\label{conclusion}

In summary, we have studied in this paper the Kitaev chains with open boundary condition by using analytic exact solution method for the non-interacting model and by using DMRG method for the interacting model.

We study a locally defined single-particle correlation function $G_{ij}$ and find that there exists a long-range edge correlation $G_{1L}$ in the topologically nontrivial phase which is absent in topologically trival phases,
while long range correlation is always absent inside bulk for all the phases. Thus, we propose that $G_{1L}$ can be used to characterize the topological order in 1+1D fermionic systems
and use it to describe quantum phase transitions between topologically trivial and nontrivial phases. It is found that $G_{1L}\propto w^z$ with $z=1$ near the critical point,
where $w=\Delta,\mu_c-\mu$, etc. is a control parameter that drives the system from a topologically nontrivial phase to a topologically trivial phase.

\section{acknowledgement}
We would like to thank Xiao-Gang Wen for helpful communications, and thank Chih-Chieh Chen for his help in DMRG programming.
This work is supported in part by National Basic Research Program of China (No.2014CB921201/2014CB921203), National Key R\&D Program of the MOST of China (No.2016YFA0300202),
NSFC (No.11374256/11774306/11674278) and the Fundamental Research Funds for the Central Universities in China. F.C.Z was also supported by the Hong Kong's University Grant Council via Grant No. AoE/P-04/08.

\appendix

\section{Exact diagonalization of non-interacting Kitaev chains with $\Delta=t$}\label{ED:non-interacting}
In this appendix, we provide details in exact diagonalization of the matrix $B$ in Eq. \eref{eq:matrixB}.
We write the matrix $B$ in the SVD form\cite{Katsura},
\begin{equation}
B=U\Lambda V^{T},
\end{equation}
where the matrix $\Lambda=\Lambda_{k}$ is diagonal. The matrices $U$ and $V$ are orthogonal transformations
\begin{eqnarray}\label{eq:UVA1}
\lambda_{k}^{1} & = & \sum_{j=1}^{L}U_{jk}\lambda_{j}^{1},\\
\lambda_{k}^{2} & = & \sum_{j=1}^{L}V_{jk}\lambda_{j}^{2},
\end{eqnarray}
which satisfy $UU^{T}=VV^{T}=\mathbf{1}$ and keep the anticommutation relations of the Majorana fermion operators
\begin{equation}
\left(\lambda_{k}^{a}\right)^{\dagger}=\lambda_{k}^{a},
\end{equation}
\begin{equation}
\left\{ \lambda_{k}^{a},\lambda_{q}^{b}\right\} =2\delta_{ab}\delta_{kq}.
\end{equation}
The energy spectra of the Hamiltonian $H_{\mu}$ are given by the singular values of the matrix $B$. We note the orthogonal matrices $U$ and $V$ diagonalize $BB^{T}$ and $B^{T}B$, respectively
\begin{eqnarray}
U^{T}BB^{T}U & = & \Lambda^{2},\\
V^{T}B^{T}BV & = & \Lambda^{2}.
\end{eqnarray}
The singular values $\Lambda_{k}$ are the non-negative square roots of the eigenvalues of $BB^{T}$. Similar diagonalization was found by Lieb et al. in the study of Heisenberg-Ising model\cite{Lieb}. The orthogonal matrices $U$ and $V$ are found to be
\begin{eqnarray}
U_{jk} & = & A_{k}\sin k\left(L+1-j\right),\\
V_{jk} & = & A_{k}\delta_{k}\sin kj,
\end{eqnarray}
where the normalization constant is
\begin{equation}
A_{k}=2\left[2L+1-\frac{\sin k\left(2L+1\right)}{\sin k}\right]^{-1/2},
\end{equation}
and
\begin{equation}
\delta_{k}=sgn \left(\frac{\sin k}{\sin kL}\right),
\end{equation}
where sgn denotes the sign function. The singular values are
\begin{equation}
\Lambda_{k}=\sqrt{\left(\mu+2t\cos k\right)^{2}+\left(2t\sin k\right)^{2}}.
\end{equation}
The $k$'s are the roots of
\begin{equation}
\frac{\sin k\left(L+1\right)}{\sin kL}=-\frac{2t}{\mu}.
\end{equation}

\begin{figure}[hptb]
\begin{center}
\includegraphics[width=8.4cm]{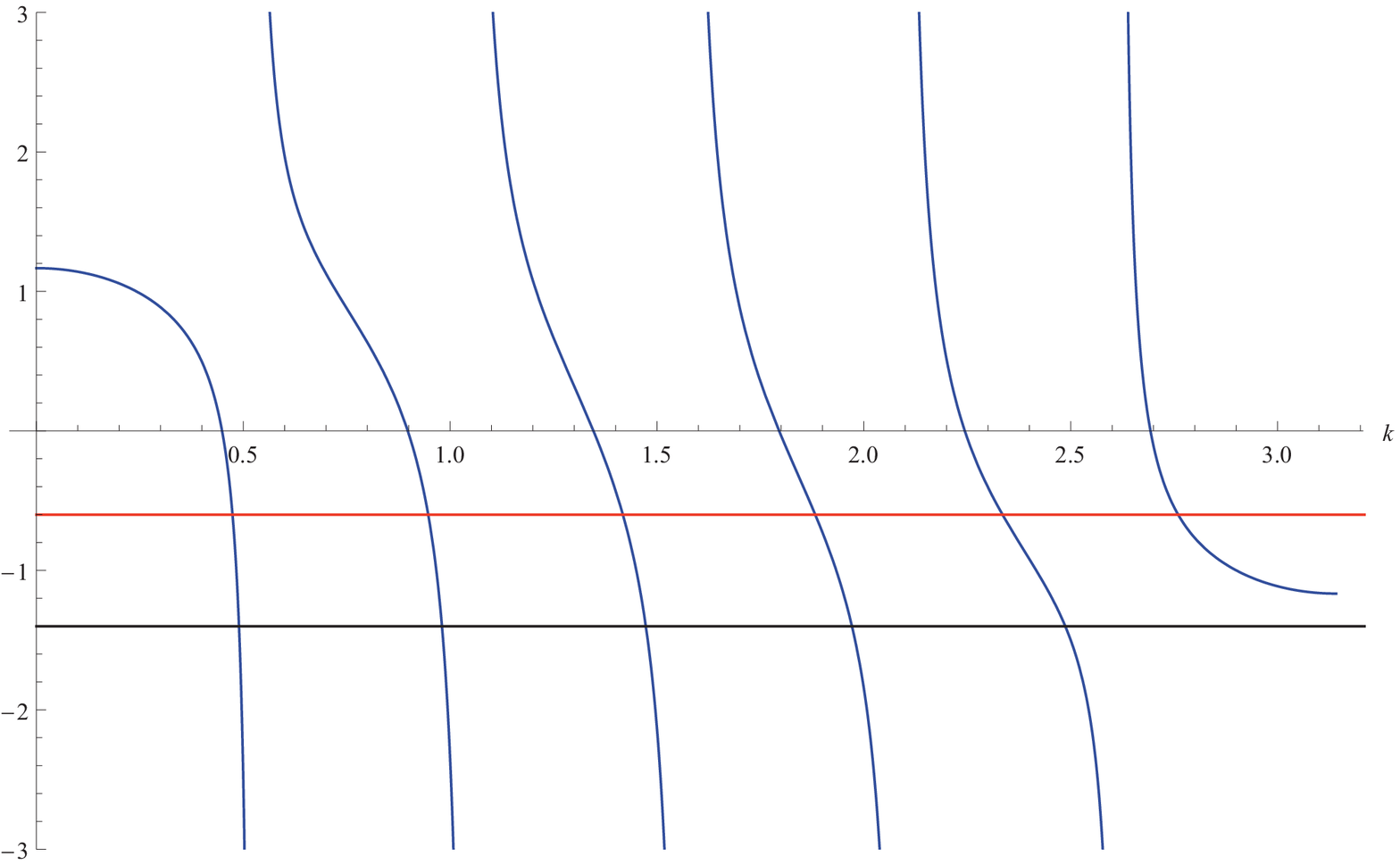}
\end{center}
\caption{
$\sin k\left(L+1\right)/\sin kL$ for $L=6$ (blue), $-2t/\mu=-0.6$
(red), $-2t/\mu=-1.4$ (black) }

\label{fig:root}
\end{figure}

The graphical solution is shown in the Fig.~\ref{fig:root}. For $\mu\geq2t$, there are $L$ real roots, including all the normal modes. For $\mu<2t$, there are $L-1$ real roots and one complex root
\begin{equation}
k_{0}=\pi+iv,
\end{equation}
with $v$ determined by
\begin{equation}\label{eq:vA}
\frac{\sinh v\left(L+1\right)}{\sinh vL}=\frac{2t}{\mu}.
\end{equation}
We consider a large open chain, i.e. $vL\gg1$
\begin{equation}
e^{v}\simeq\frac{2t}{\mu}-\left(\frac{2t}{\mu}-\frac{\mu}{2t}\right)\left(\frac{\mu}{2t}\right)^{2L}.
\end{equation}
Then for this special mode we have
\begin{eqnarray}
U_{jk_{0}} & = & A_{k_{0}}\sinh v\left(L+1-j\right),\\
V_{jk_{0}} & = & A_{k_{0}}\sinh v j,
\end{eqnarray}
the normalization constant becomes
\begin{equation}
A_{k_{0}}=2e^{- v L}\left(1-e^{-2v}\right)^{1/2},
\end{equation}
and the singular value is
\begin{equation}
\Lambda_{k_{0}}=\Bigl(\frac{2t}{\mu}-\frac{\mu}{2t}\Bigr)\Bigl(\frac{\mu}{2t}\Bigr)^{L}.
\end{equation}

\end{document}